\documentclass{PoS}

\title{Low temperature condensation and scattering data}

\ShortTitle{Low temperature condensation and scattering data}

\author{\speaker{Oliver Orasch}\\
        University of Graz, Institute of Physics, 8010 Graz, Austria\\
        E-mail: \email{oliver.orasch@uni-graz.at}}
        
\author{Christof Gattringer\\
        University of Graz, Institute of Physics, 8010 Graz, Austria\\
        E-mail: \email{christof.gattringer@uni-graz.at}}
        
\author{Mario Giuliani\\
        University of Graz, Institute of Physics, 8010 Graz, Austria\\
        E-mail: \email{mario.giuliani@uni-graz.at}}

\abstract{We study $\phi^4$ lattice field theory at finite chemical potential $\mu$ in two and four dimensions, 
using a worldline representation that overcomes the complex action problem. We compute the particle number 
at very low temperature as a function of $\mu$ and determine the first three condensation thresholds, where 
the system condenses 1, 2 and 3 particles. The corresponding critical values of the chemical potential can be 
related to the 1-, 2- and 3-particle energies of the system, and we check this relation with a direct spectroscopy
determination of the $n$-particle energies from $2n$-point functions. We analyze the thresholds as a function of the 
spatial size of the system and use the known finite volume results for the $n$-particle energies to relate the thresholds 
to scattering data. For four dimensions we determine the scattering length from the 2-particle threshold, while in 
two dimensions the full scattering phase shift can be determined. In both cases the scattering data computed
from the 2-particle threshold already allow one to determine the 3-particle energy. In both, two and four dimensions 
we find very good agreement of this ''prediction'' with direct determinations of the 3-particle energy from either 
the thresholds or the 6-point functions. The results show that low temperature condensation is indeed governed 
by scattering data.}

\FullConference{The 36th Annual International Symposium on Lattice Field Theory - LATTICE2018\\
		22-28 July, 2018\\
		Michigan State University, East Lansing, Michigan, USA.}

\begin{document}

\section{Introduction}
In recent years considerable progress was made with overcoming the complex action problem at finite 
density for several lattice field theories. It was possible to exactly map the partition sum 
of these systems to a representation in terms of worldlines (and/or worldsheets) where all contributions to the 
partition sum are real and positive, such that a Monte Carlo simulation can be done directly in terms of the 
worldlines (see, e.g., \cite{Endres,Weisz,phi4_1,phi4_2} for work on the $\phi^4$ theory studied here).
With the worldline approach it is possible to address new physics questions related to finite density. 
An example is condensation of particles at low temperatures, which is the topic of this contribution. 

\begin{figure}[b!]
	\centering      
	\hspace*{-3mm}
	\includegraphics[width=90mm,clip]{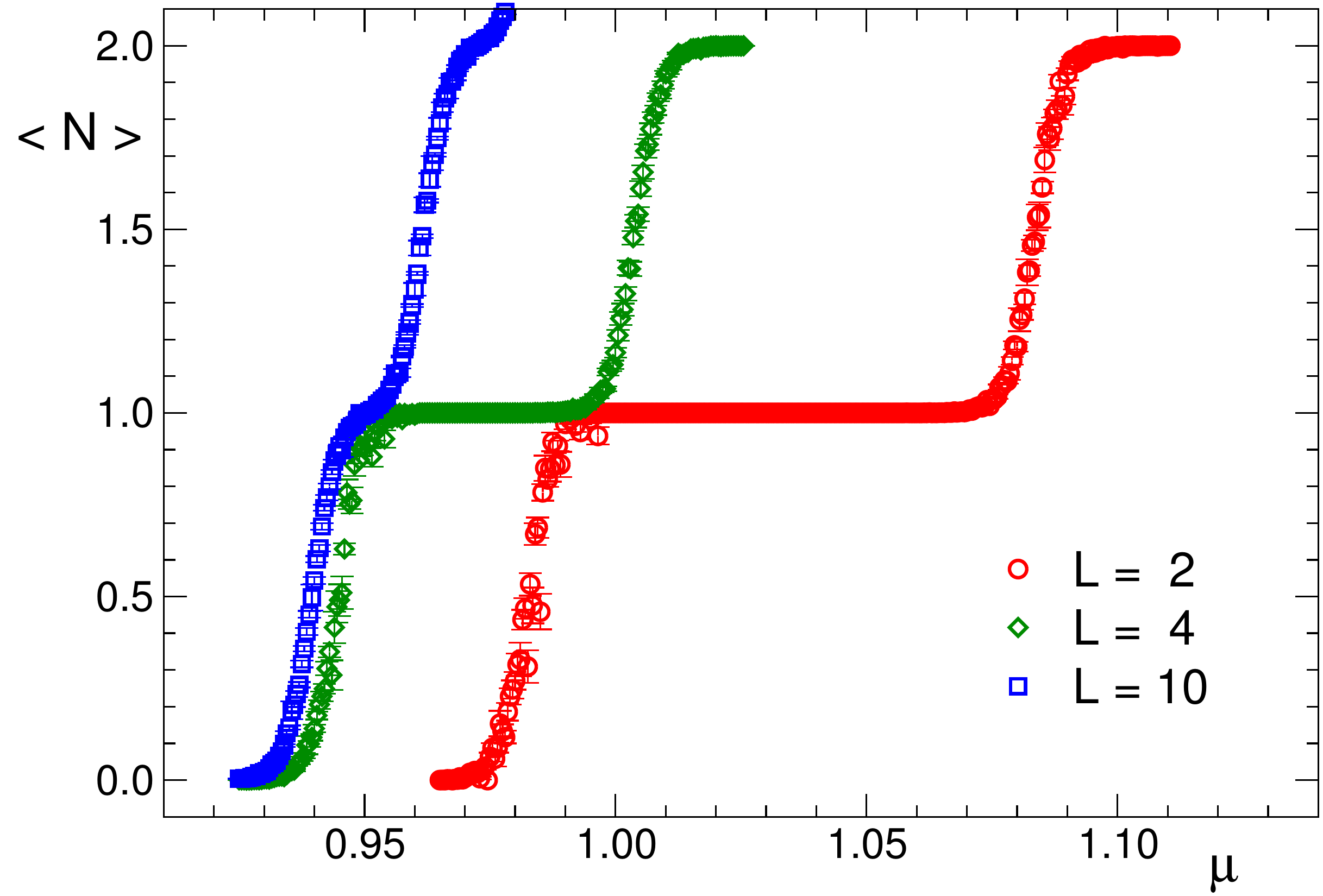}
	\caption{The expectation value of the particle number $\langle N \rangle$ as a function of the chemical 
	potential $\mu$ (in lattice units). We show the results for the 2d case at $N_t = 400$ and different values of $L$.}
	\label{condexample}
\end{figure}

To illustrate the condensation phenomenon we study here, in Fig.~\ref{condexample} we show the results 
for the expectation value of the particle number $\langle N \rangle$ versus the chemical potential $\mu$. 
The results are for $\phi^4$ theory in 2d on a $L \times N_t$ lattice with $N_t = 400$ and three different 
values of $L$. The temperature is very low ($T = 1/N_t = 0.0025$ in lattice units) and indeed we observe 
condensation as a function of $\mu$ (details see below). For each of the three values of $L$ 
the particle number $\langle N \rangle$ quickly rises from $\langle N \rangle = 0$ to $\langle N \rangle = 1$ 
at some critical chemical potential value $\mu_1(L)$, then further to $\langle N \rangle = 2$ 
at a second critical value $\mu_2(L)$ and similar for higher particle number sectors. Note that at zero 
temperature ($N_t = \infty$) one expects discontinuous jumps of $\langle N \rangle$ which are here rounded 
by temperature effects. Nevertheless we can identify the critical values $\mu_n(L)$ and determine their values 
as a function of $L$. The values $\mu_n(L)$ correspond to the values of the chemical potential where we observe 
condensation of another particle visible in the step from $\langle N \rangle = n-1$ to $\langle N \rangle = n$.

In \cite{Bruckmann} it was shown that at very low temperature 
the condensation thresholds $\mu_n$ are related to the physical mass 
$m(L)$ and the $n$-particle energies $W_n(L)$ via the relations
\begin{equation}
 m(L) \; = \; \mu_1(L) \; , \qquad W_n(L) \; = \; \sum_{k=1}^n \mu_k(L) \; ,
 \label{mu_energy}
 \end{equation}
where we now made explicit, that not only the $\mu_n(L)$, but also the physical mass $m(L)$ and the 
$n$-particle energies $W_n(L)$ depend on the spatial extent $L$. It has been known since the pioneering 
paper \cite{huang_yang} that the dependence of the 2- and 3-particle energies $W_2(L)$ and $W_3(L)$ on the
spatial extent $L$ can be parameterized in terms of scattering data of the underlying theory. Thus we conclude from 
(\ref{mu_energy}) that the condensation thresholds $\mu_n(L)$ are governed by the scattering data.

Demonstrating and analyzing the connection between low temperature condensation and scattering data in 
$\phi^4$ theory in two and four dimensions is the topic of this contribution (see also \cite{prl}).

\section{Worldline representation and Monte Carlo simulation}

The system where we explore the relation between condensation and scattering data is the complex $\phi^4$ field in
$d = 2$ and $d = 4$ dimensions. The lattice action is given by 
\begin{equation}
S[\phi]  \; = \;   
\sum_{x \in \Lambda} \bigg( \eta \, |\phi _{x}|^2 \; + \; \lambda \, |\phi _{x}|^4  
\;  -  \; \sum_{\nu = 1}^{d} 
\left[ e^{\,\mu \delta_{\nu,d}}\phi _{x}^{\ast} \phi _{x+\hat{\nu}} + 
e^{\,-\mu \delta_{\nu,d}}\phi _{x+\hat{\nu}}^{\ast} \phi _{x} \right]\bigg) \; ,
\label{action_conventional}
\end{equation}
where $\eta \equiv 2d + m^{2}_b$ with $m_b$ the bare mass parameter. $\lambda$ is the quartic coupling and $\mu$ 
the chemical potential. The fields $\phi_x$ are assigned to the sites $x$ of a lattice of size $L^{d-1} \times N_t$.

At $\mu \neq 0$ the action is complex and the Boltzmann factor $e^{-S[\phi]}$ cannot be used as a probability 
in a Monte Carlo simulation. This complex action problem of the conventional representation (\ref{action_conventional}) 
can be solved by exactly mapping the system to a worldline representation (see, e.g., \cite{phi4_1} for a derivation of
the form we use here). In the worldline representation the partition sum reads ($\beta \equiv N_t$)
\begin{equation}
Z   \; = \; \sum_{\{k\}}  \left[ \prod_{x} \delta \left(\vec{\nabla} \cdot \vec{k}_{x} \right) \right] \, 
e^{\, \mu \, \beta \, \omega[k]} \; B[k] \; .
\label{worldlineZ}
\end{equation}
$Z$ is a sum over configurations of the worldline variables $k_{x,\nu} \in \mathbb{Z}$
assigned to the links of the lattice. They have to obey constraints which have the form of a product 
over Kronecker deltas $\delta(j) \equiv \delta_{j,0}$ at all sites $x$. At each $x$ the Kronecker deltas 
enforce $\vec{\nabla} \cdot \vec{k}_{x}  \equiv   \sum_{\nu}(k_{x,\nu} - k_{x-\hat{\nu},\nu}) = 0$, i.e., 
zero divergence for $k_{x,\nu}$, and as a consequence the 
worldline variables $k_{x,\nu}$ must form closed loops of conserved flux. By $\omega[k]$ we denote the 
total winding number of the $k$-flux around the compact time direction and the 
chemical potential couples to $\omega[k]$
in the form $e^{\mu \beta \omega[k]}$. The observable we
need for our analysis is the expectation value of the particle number $\langle N \rangle = \partial \ln Z / \partial \beta \mu = 
\langle \omega[k] \rangle_{wl}$, where $\langle .. \rangle_{wl}$ denotes the vacuum expectation in the 
worldline representation.  

The configurations of the worldline variables $k_{x,\nu}$ come with a real and positive weight factor 
\begin{equation}
B[k] \; = \; \sum_{\{a\}} \, \prod_{x,\nu}\frac{1}{(a_{x,\nu}+|k_{x,\nu}|)! \, a_{x,\nu}!} \,  \prod_{x} I(s_x) \; \; \quad
\mbox{with} \quad \; \; I(s_x) \; = \; \int_{0}^{\infty} \!\! d r \; r^{\, s_x + 1} \, e^{\, -\eta \, r^2 \, - \, \lambda \, r^4} \; .  
\label{weights} 
\end{equation}
$B[k]$ is a sum over configurations $\sum_{\{a\}}$ of auxiliary link variables $a_{x,\nu} \in \mathbb{N}_0$, and by  
$s_x$  we denote the non-negative integer combination 
$s_x \; = \; \sum_{\nu}\big[|k_{x,\nu}| +  |k_{x-\hat{\nu}}| + 2(a_{x,\nu} +  a_{x-\hat{\nu}})\big] $, which appears 
as an argument in the integrals $I(s_x)$ that come from integrating out the radial degrees of freedom of the 
original field variables at site $x$. They are pre-calculated and stored for the simulations. 

All weight factors in the worldline representation are real and positive such that the complex action problem is solved. 
Concerning the details of the updates for the
worldline variables $k_{x,\nu}$ and the auxiliary variables $a_{x,\nu}$ we refer to \cite{algorithm,preliminary}.
In 4d we use lattices with $N_t = 320$ and 640, and $L$ between 3 and 10 at coupling values of 
$\eta = 7.44$ and $\lambda = 1.0$, with a statistics of $2 \times 10^5$ configurations. 
In 2d the corresponding parameters are $N_t = 400$, $L$ between 2 and 16 with $\eta = 2.6$, $\lambda = 1.0$ 
and a statistics of $4 \times 10^5$.

\section{Analysis of the 4d case}

After computing $\langle N \rangle$ as a function of $\mu$ we identify the steps where
$\langle N \rangle$ transits from $\langle N \rangle = n-1$ to $\langle N \rangle = n$ (compare Fig.~\ref{condexample}). 
To determine the corresponding critical values $\mu_n$ we fit the data for $\langle N \rangle$ in the vicinity
of the steps with the logistic function $\langle N \rangle \!=\! [1\!\,+\!\, \exp(-a_{n}[\mu\!-\!\mu_n])]^{-1}\!+\!n\!-\!1$.
Using (\ref{mu_energy}) we then compute $m(L)$, $W_2(L)$ and $W_3(L)$ from the critical values $\mu_n(L)$. 

\begin{figure}[t!]
	\centering      
	\hspace*{-3mm}
	\includegraphics[width=80mm,clip]{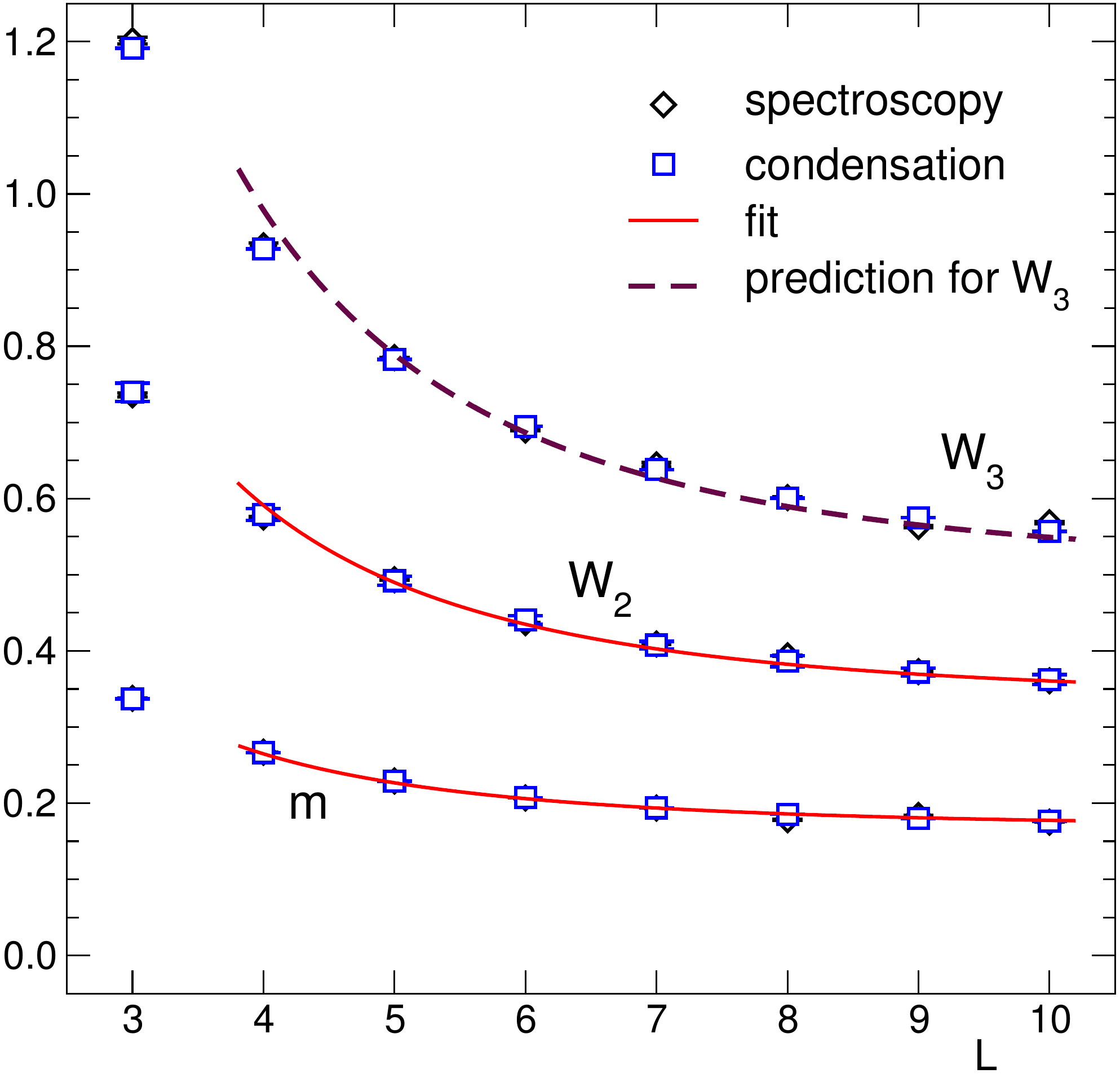}
	\caption{The physical mass $m(L)$ and the 2- and 3-particle energies $W_2(L)$ and $W_3(L)$ for the 
	4d case as a function of the lattice extent $L$ (figure from \cite{prl}). 
	We show the results determined from the condensation steps (blue squares) 
	and compare them to the results from spectroscopy (black diamonds). The full red curves are the fits of $m$ and 
	$W_2$ with (\ref{mL}) and (\ref{W2}). The dashed maroon curve is the function $W_3$ from Eq.~(\ref{W3})
	when using the scattering length $a$ from the fit of $W_2$ as input.}
	\label{energy_plot}
\end{figure}

In Fig.~\ref{energy_plot} we show the results for $m(L)$, $W_2(L)$ and $W_3(L)$ determined from the critical 
chemical potential values $\mu_n(L)$ as squares. To test the relations (\ref{mu_energy}) and the reliability of our 
determination of the critical values $\mu_n(L)$, we computed $m(L)$, $W_2(L)$ and $W_3(L)$ also in a spectroscopy 
analysis based on $2n$-point functions calculated at $\mu = 0$ in the conventional representation 
(\ref{action_conventional}). The corresponding results are shown as diamonds in Fig.~\ref{energy_plot} and 
coincide almost perfectly with the data from the condensation steps. This cross check confirms the interpretation 
of the critical chemical potential values as combinations of multi-particle energies.

The next step is to invoke the finite volume relations for $m(L)$ \cite{kari}, the result \cite{huang_yang,Luscher_w2} 
for the 2-particle energy $W_2(L)$ (using the notation of \cite{sharpe}) and the results 
\cite{beane,sharpe,sharpe1,sharpe2,sharpe3} for the 3-particle energy $W_3(L)$ 
(the numerical constants ${\cal I}$ and ${\cal J}$ are given by ${\cal I} = -8.914, {\cal J} = 16.532$):
\begin{eqnarray}
m(L) & = & m_\infty + \frac{A}{L^{\frac{3}{2}}} \, e^{ - L \; m_\infty } ,
\label{mL} \\
W_2(L) & = & 2m  +  \frac{4\pi a}{m L^3} \! 
\Bigg[ 1 - \frac{a}{L} \frac{{\cal I}}{\pi} 
 + 
\bigg(\!\frac{a}{L}\! \bigg)^{\!\!2} \, \frac{ {\cal I}^{\,2} \!-\! {\cal J}}{\pi^2}  + 
{\cal O} \! \bigg(\!\frac{a}{L}\! \bigg)^{\!\!3}\Bigg]\! ,
 \label{W2} \\
 W_3(L) & = & 3m +  \frac{12\pi a}{m L^3} \! 
\Bigg[ 1 - \frac{a}{L} \frac{{\cal I}}{\pi}  + 
\bigg(\!\frac{a}{L}\! \bigg)^{\!\!2} \, \frac{ {\cal I}^{\,2} \!+ \!{\cal J}}{\pi^2}  +  
{\cal O} \! \bigg(\!\frac{a}{L}\! \bigg)^{\!\!3}\Bigg]\! .
\label{W3}
\end{eqnarray}
Up to order $1/L^5$ only three parameters are needed to describe the data: the infinite volume mass 
$m_\infty$, the amplitude $A$ and the scattering length $a$. Fitting the data for $m(L)$ with the relation (\ref{mL}) 
we find a value of $m_\infty = 0.168(1)$ in lattice units. For fitting $W_2(L)$ we use (\ref{W2}), with the mass parameter
$m$ on the rhs.~replaced by the corresponding values $m(L)$. This gives rise to a value of $a = - 0.078(7)$ for the 
scattering length in lattice units and a value of $a \, m_\infty = - 0.013(1)$ for the dimensionless product of
$a$ and $m_\infty$. The functions (\ref{mL}) and (\ref{W2}) with the fit values for $m_\infty$, $A$ and 
$a$ are shown as full red curves in Fig.~\ref{energy_plot} and describe the data for $m(L)$ and $W_2(L)$ 
very well (with the exception of the smallest $L$ where higher corrections in $1/L$ would be necessary). 

Having determined the mass and the scattering length, no further parameters are necessary to describe 
$W_3(L)$ with (\ref{W3}). Inserting the fit value for $a$ and again using $m(L)$ in the rhs.\ of (\ref{W3}),
we thus get a ''prediction'' for the data $W_3(L)$. This prediction is shown as a dashed 
curve in Fig.~\ref{energy_plot} and obviously describes the data for $W_3(L)$ very well (again with the exception of the 
smallest $L$).  

This concludes the discussion of the 4d case and our results confirm the relations (\ref{mu_energy}) of the 
condensation thresholds to multi-particle energies, which in turn are described by scattering data. Thus we have 
quantitatively established the connection of condensation and scattering data.

\section{Analysis of the 2d case}

Also in the 2d case we determined the critical values $\mu_n(L)$ from fitting the steps of $\langle N \rangle$ and then 
computed $m(L)$, $W_2(L)$ and $W_3(L)$ using the relations (\ref{mu_energy}). We cross-checked these results with
a spectroscopy calculation in the conventional representation and again found very good agreement between the 
condensation and the spectroscopy results.

\begin{figure}[t!]
	\centering      
	\hspace*{-3mm}
	\includegraphics[height=62mm,clip]{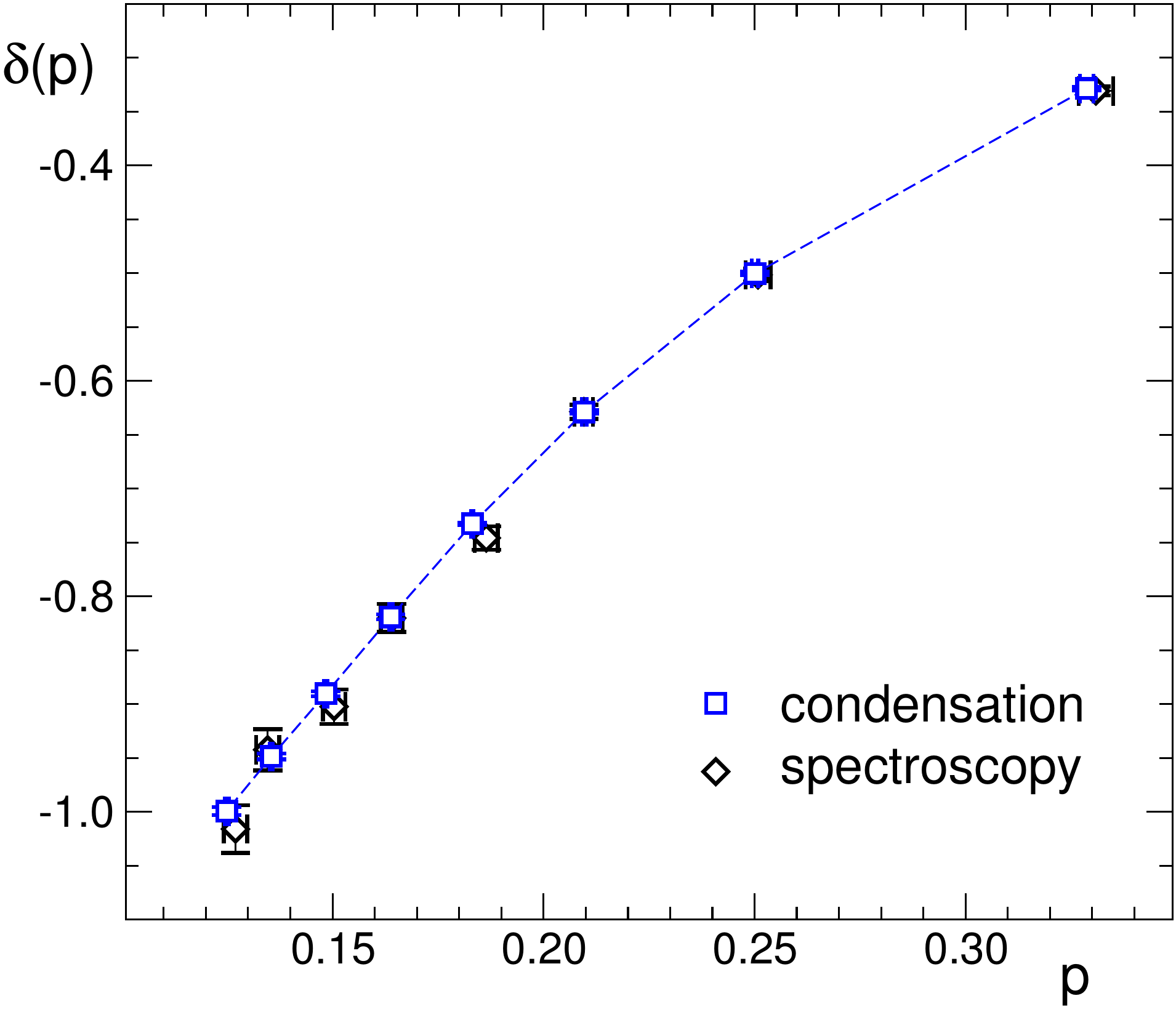}
	\hspace{4mm}
	\includegraphics[height=62mm,clip]{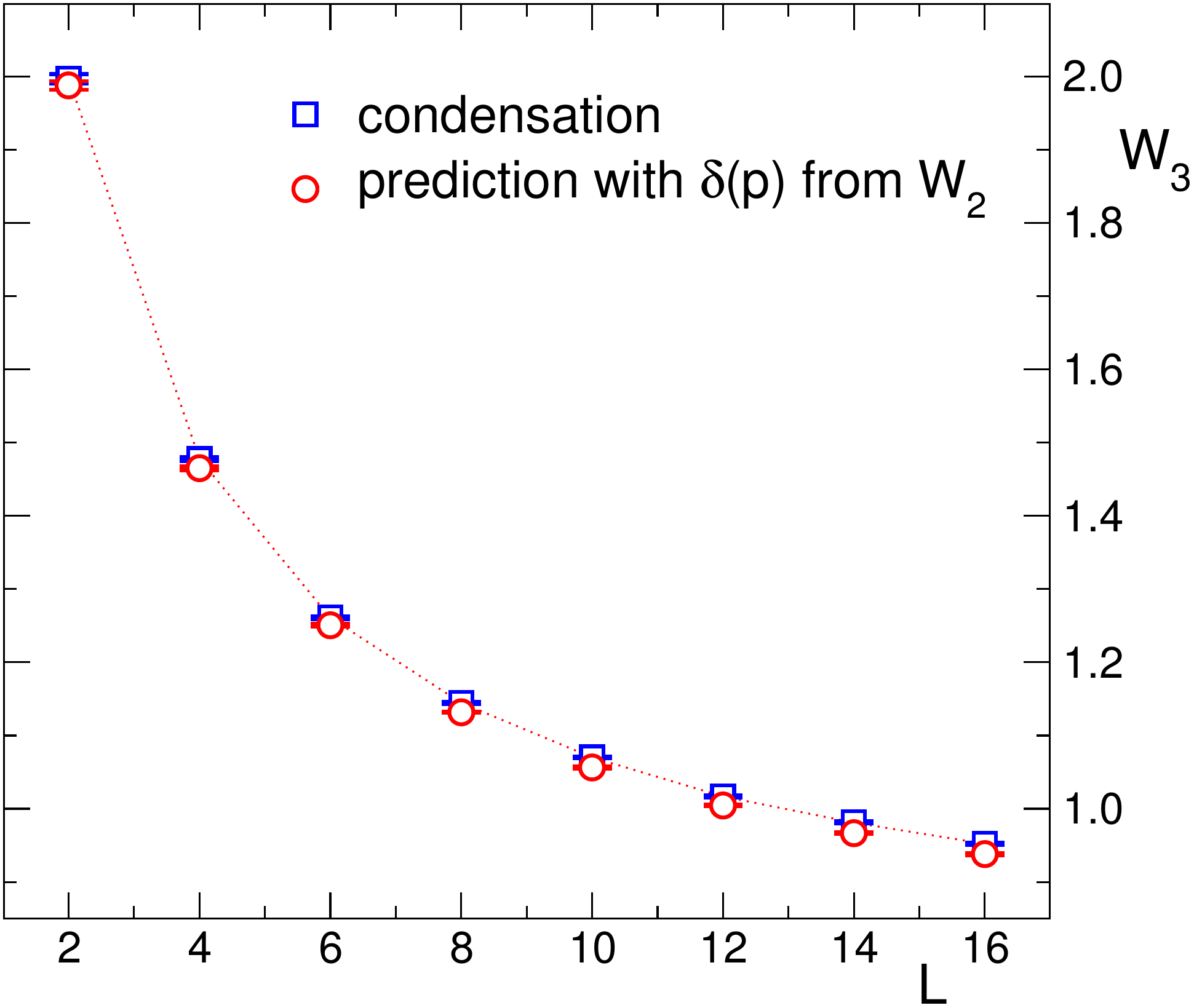}
	\caption{Lhs.: The scattering phase shift $\delta(p)$ versus $p$. We compare the results from the 
	condensation threshold (blue squares) to reference data from spectroscopy (black diamonds).
	Rhs.: The 3-particle energy $W_3$ as a function of $L$. We show the results of the direct determination 
        from the condensation thresholds (blue squares) and compare it to a prediction based on 
        $\delta(p)$ determined from $W_2$.}
	\label{plot2D}
\end{figure}

The next step is the finite volume analysis of $m(L)$, $W_2(L)$ and $W_3(L)$. As before the mass $m(L)$ can be 
described with a 2-parameter ansatz, which in 2d reads $m(L) = m_\infty + A \, e^{-m_\infty \, L} / \sqrt{L}\, $. 
For analyzing the 2-particle energy $W_2(L)$ we follow the approach \cite{Luscher_2d} that is applicable to 
short range potentials. Outside the interaction range the wave function is a 2-particle plane wave 
$\psi = e^{- i x_1 p_1} \, e^{- i x_2 p_2}$ with momenta $p_1$ and $p_2$. The corresponding energy is 
$W_2(L) =  \sum_{j=1}^2 \sqrt{ m(L)^2 + p_j^2}$. We rewrite the wave function $\psi$ by using the center 
of mass coordinate $(x_1 + x_2)/2$ and the relative coordinate $r = x_1 - x_2$.  The energy values $W_2(L)$ 
determined from the condensation steps correspond to vanishing total momentum $p_1 + p_2 = 0$, and we set 
$p_1 = p = -p_2$.  For vanishing total momentum the wave function then has the form $\psi = e^{-i p r}$ and the 
2-particle energy is given by 
\begin{equation}
W_2(L) \; = \; 2 \sqrt{ m(L)^2 + p^2} \; .
\label{2dW2} 
\end{equation}
For finalizing the connection between the 2-particle energy and the scattering data we need to invoke the quantization 
for the momenta $p$ in a finite box of size $L$. This condition is obtained from the boundary condition for $\psi$ which 
connects the wave function at $r = 0$ to its value at $r = L$ and reads $e^{-i \,p L} = e^{\, i \,2 \delta(p)}$. It expresses 
the fact that the plane wave solution is correct only outside the interaction range, and that when connecting $r = 0$ with
$r = L$ one has to take into account the phase shift $\delta(p)$ that is picked up when the two particles interact. Thus 
we obtain 
\begin{equation}
\delta(p) \; = \; - \frac{p L}{2} \; .
\label{2ddelta}
\end{equation}
The two equations (\ref{2dW2}) and (\ref{2ddelta}) constitute the connection between scattering data and $W_2(L)$. 
We can use the numerically determined values for $W_2(L)$ to compute from (\ref{2dW2}) the corresponding relative 
momenta $p$, and then use these to compute the scattering phase shift $\delta(p)$. We show the corresponding results 
in the lhs.\ plot of Fig.~\ref{plot2D} and compare the data from a determination based on the condensation thresholds to 
those from a determination based on standard spectroscopy. 

Before we discuss $W_3(L)$, we stress that at fixed couplings the phase shift is only 
a function of the lattice size $L$: the momentum $p$ determined from (\ref{2dW2}) depends only on the 
lattice size $L$, i.e., we have $p = p(L)$. Thus the phase shift  from (\ref{2ddelta}) is given by 
$\delta(p(L)) =  - \frac{p(L) \, L}{2}  \equiv \delta(L)$.

Similar to the 4d case we now use the scattering data determined from $W_2(L)$ to ''predict'' $W_3(L)$ and 
thus the third critical chemical potential value $\mu_3(L)$. The approach is a generalization of the strategy
\cite{Luscher_2d} we have followed for analyzing $W_2(L)$. Again we make a plane wave ansatz   
$\psi = e^{- i x_1 p_1} \, e^{- i x_2 p_2} \, e^{- i x_3 p_3}$ for three particles which describes the system when 
all three particles are sufficiently remote from each other. The corresponding energy is 
$W_3(L) =  \sum_{j=1}^3 \sqrt{ m(L)^2 + p_j^2}$. 

As before we introduce the center of mass coordinate $(x_1 + x_2 + x_3)/3$, as well as the relative coordinates 
$r_{2} = x_2 - x_1$ and $r_{3} = x_3 - x_1$. Using these to parameterize $\psi$ and demanding total vanishing 
momentum $p_1 + p_2 + p_3 = 0$, we find $\psi = e^{- i r_2 p_2} \, e^{- i r_3 p_3}$ and $p_1 = - p_2 - p_3$. 
This 3-particle wave function has to obey two quantization conditions of the form (\ref{2ddelta}) that contain 
$p_2$ and $p_3$. Using the fact that the phase shift is only a function of $L$ we can determine $p_2$ and $p_3$ 
as $p_2 = p_3 = - 2 \delta(L) / L$. Inserting these values and $p_1 = - p_2 - p_3$ into 
$W_3(L) =  \sum_{j=1}^3 \sqrt{ m(L)^2 + p_j^2}$ we obtain our prediction for $W_3(L)$. The corresponding 
values are shown as red circles in the rhs.\ plot of Fig.~\ref{plot2D}. We compare them to the results of a direct 
determination from all three condensation thresholds. The results agree very well and we 
conclude that the structure of the condensation thresholds $\mu_n(L)$ can indeed be correctly described 
with the scattering data of the theory.

We remark that the 2- and 3-particle energies $W_2(L)$ and $W_3(L)$ can be analyzed 
with a different approach \cite{Guo1,Guo2}, where one uses the exact solution for the scattering phase shift 
which depends on a single parameter, the amplitude $V_0$ of the point-like interaction. This parameter can be 
determined from $W_2(L)$ and subsequently used for $\delta$ in the 3-particle quantization conditions to determine 
the two independent momenta $p_2$ and $p_3$ needed to compute $W_3(L)$.

\section{Concluding remarks}

In this contribution we have shown for a simple scalar field theory in two and four dimensions that 
low temperature condensation is governed by the scattering data of the theory. This relation is expected 
to be a general non-perturbative feature, but in order to study it on the lattice usually a complex action 
problem has to be solved, which so far has been achieved for only a few systems. However, there exist 
interesting theories which are already free of the complex action problem. Examples are lattice field theories 
based on the gauge group SU(2), and more interestingly, QCD with isospin chemical potential where the 
condensation of pions is expected to be related to pion scattering data. For these systems an analysis along 
the lines sketched here should be possible.

\vskip3mm
\noindent
{\bf Acknowledgements:} We thank F.\ Bruckmann, P.\ Guo, M. Hansen, T.\ Kloiber, A.\ Maas, C.B.\ Lang, C.\ Marchis, 
S.\ Sharpe and T.\ Sulejmanpasic for discussions. This work is supported by the Austrian Science 
Fund FWF, grant I 2886-N27 and the FWF DK W 1203, ''Hadrons in Vacuum, Nuclei and Stars".

\end{document}